\documentclass[twocolumn]{aastex631}
\usepackage{hyperref}
\usepackage{epstopdf}
\usepackage{amsmath}
\usepackage{multirow}
\usepackage{soul}

\setulcolor{red}
\setstcolor{red}

\newcommand{\sixteenhundred}{$1600$\,{\AA}}
\newcommand{\threezerofour}{$304$\,{\AA}}
\newcommand{\oneseventyone}{$171$\,{\AA}}
\newcommand{\oneninetythree}{$193$\,{\AA}}
\newcommand{\twooneone}{$211$\,{\AA}}
\newcommand{\threethirtyfive}{$335$\,{\AA}}
\newcommand{\onethirtyone}{$131$\,{\AA}}

\begin{document}

\title{Exploring the dynamic rotational profile of the hotter solar atmosphere: A multi-wavelength approach using SDO/AIA data}
\shorttitle{Rotational profile of the solar atmosphere}
\shortauthors{S. Routh et al.}

\correspondingauthor{Dipankar Banerjee}
\email{dipu@aries.res.in}

\author[0009-0008-5834-4590]{Srinjana Routh}
\affiliation{Aryabhatta Research Institute of Observational Sciences, Nainital-263002, Uttarakhand, India}
\affiliation{Mahatma Jyotiba Phule Rohilkhand University, Bareilly-243006, Uttar Pradesh, India }

\author[0000-0003-3191-4625]{Bibhuti Kumar Jha}
\affiliation{Southwest Research Institute, Boulder, CO 80302, USA}

\author[0009-0003-1377-0653]{Dibya Kirti Mishra}
\affiliation{Aryabhatta Research Institute of Observational Sciences, Nainital-263002, Uttarakhand, India}
\affiliation{Mahatma Jyotiba Phule Rohilkhand University, Bareilly-243006, Uttar Pradesh, India }

\author[0000-0001-9628-4113]{Tom Van Doorsselaere}
\affiliation{Centre for mathematical Plasma Astrophysics, Mathematics Department, KU Leuven, Celestijnenlaan 200B \\ bus 2400, B-3001 Leuven, Belgium}

\author[0000-0002-6954-2276]{Vaibhav Pant}
\affiliation{Aryabhatta Research Institute of Observational Sciences, Nainital-263002, Uttarakhand, India}

\author[0000-0002-5014-7022]{Subhamoy Chatterjee}
\affiliation{Southwest Research Institute, Boulder, CO 80302, USA}

\author[0000-0003-4653-6823]{Dipankar Banerjee}
\affiliation{Aryabhatta Research Institute of Observational Sciences, Nainital-263002, Uttarakhand, India}
\affiliation{Indian Institute of Astrophysics, Koramangala, Bangalore 560034, India}
\affiliation{Center of Excellence in Space Sciences India, IISER Kolkata, Mohanpur 741246, West Bengal, India}

\begin{abstract}
Understanding the global rotational profile of the solar atmosphere and its variation is fundamental to uncovering a comprehensive understanding of the dynamics of the solar magnetic field and the extent of coupling between different layers of the Sun. In this study, we employ the method of image correlation to analyze the extensive dataset provided by the Atmospheric Imaging Assembly of the Solar Dynamic Observatory in different wavelength channels. We find a significant increase in the equatorial rotational rate ($A$) and a decrease in absolute latitudinal gradient ($|B|$) at all temperatures representative of the solar atmosphere, implying an equatorial rotation up to $4.18\%$  and $1.92\%$ faster and less differential when compared to the rotation rates for the underlying photosphere derived from Doppler measurement and sunspots respectively. In addition, we also find a significant increase in equatorial rotation rate ($A$) and a decrease in differential nature ($|B|$ decreases) at different layers of the solar atmosphere. We also explore a possible connection from the solar interior to the atmosphere and interestingly found that $A$ at $r=0.94\,\mathrm{R}_{\odot}, 0.965\,\mathrm{R}_{\odot}$ show an excellent match with \oneseventyone, \threezerofour\, and \sixteenhundred, respectively. Furthermore, we observe a positive correlation between the rotational parameters measured from \sixteenhundred, \onethirtyone, \oneninetythree\, and \twooneone\, with the yearly averaged sunspot number, suggesting a potential dependence of the solar rotation on the appearance of magnetic structures related to the solar cycle or the presence of cycle dependence of solar rotation in the solar atmosphere.
\end{abstract}


\section{Introduction} \label{sec:intro}
Rotation is a fundamental aspect in the pursuit of a comprehensive understanding of our nearest star, the Sun. The study of solar rotation has been a persistent topic in solar physics since the 17th century and has become increasingly important in recent years due to its strong connection with the solar magnetic field \citep{Parker1955,Parker1955a,Charbonneau2010}. Early studies on the differential rotation in the photosphere of the Sun relied on tracking of prominent photospheric magnetic features called sunspots \citep{Carrington1859,Newton1951}, which allowed for the measurement of photospheric differential rotation in the form of \citep{WeberThesis},
\begin{equation}
\Omega= A + B\sin^2{\theta} + C \sin^4{\theta}, 
\label{diffequation}
\end{equation}
where $\theta$ is the latitude, $A$ is the equatorial rotation rate, and $B$ and $C$ are the coefficients of a quadratic expansion in $\sin^2{\theta}$, often physically interpreted as latitudinal gradients \citep{Li2013Solar-cycle-related}.

In the past century, advances in measuring techniques and instruments have significantly improved the accuracy of sunspot tracking \citep{Ward1966,Balthasar1986,gupta1999,javaraiah2005,Jha2021, jha2022thesis}, and have also led to the development of new measurement techniques such as spectroscopy \citep{Howard1970, Howard1984, Snodgrass1984,Snodgrass1990,Vats2001}. Furthermore, extensive research, in conjunction with the more recent field of helioseismology \citep{Antia1998,Komm2008,Howe2009}, has enabled us to gain a comprehensive understanding of the rotational profile of the Sun till photosphere, including its variations with depth. However, a complete understanding of the rotational profile of the Sun above the photosphere and its variation with temperature (or height) remains elusive.

Initial investigations into the rotational profile of the higher solar atmosphere, where magnetic field dominates the dynamics \citep{Stix1976,Gary2001,Gomez2019}, suggested a faster rate of rotation than the photosphere \citep{Hale1908,Evershed1925, Aslanov1964, Hansen1968, Livingston1969}. These findings were contradicted by the subsequent studies that suggested a rotational profile of different parts of the solar atmosphere to be similar to that of the photosphere or sunspots, if not even slower \citep{Fisher1984,Brajsa1999,Brajsa2004,Bertello2020}.  Studies also made efforts to utilise higher atmospheric features like filaments \citep{Glackin1974,Brajsa1991,Japaridze1992}; coronal bright points \citep[CBPs;][]{Brajsa2004, Sudar2015}; coronal streamers \citep{Morgan2011,Edwards2022}; magnetic loops \citep[e.g.,][]{Pneuman1971}; coronograph images \citep[e.g.,][]{Lewis1999,Mancuso2020}; Ca$^{+}$ network and plages \citep[e.g.,][]{Schroeter1978,Bertello2020,Mishra2024}; soft X-ray observation \citep[SXR; ][]{Chandra2010}, wavelength bands like \oneseventyone, \oneninetythree, \threezerofour, etc. \citep{Sharma2020,Sharma2021} and radio flux information \citep[e.g.,][]{Vats2001,Bhatt2017} to obtain the rotational profile of the different layers of hotter solar atmosphere.  Additionally, some studies reported an increase in rotation rate with temperature/height \citep[e.g.,][]{AdamsTang1977,Vats2001,Sharma2020}, while others found contrary results \citep[e.g.,][]{Bhatt2017,Badalyan2018}. These studies used various methods, including the tracer method \citep[e.g,][]{Schroeter1978}, periodogram \citep[e.g,][]{Weber1999}, auto-correlation\citep[e.g.,][]{Sharma2020,Sharma2021} and cross-correlation method \citep[e.g.,][]{Bertello2020,Mishra2024} to obtain diverse results, which have been unable to resolve the problem of atmospheric solar rotation, persisted for over a century.

Studies akin to that of \cite{Glackin1974,ternullo_rotation_1986,Japaridze1992,Komm1993torsional} have linked the cause of these reported differences in the obtained rotational profiles based on the nature of the tracer chosen. However, other studies, such as those by \cite{Altrock2003,Mishra2024} attributed the probable cause to the characteristics of the data selected for analysis. Additionally, several studies have also explored the correlation between the solar cycle and the rotational profile of the solar corona and transition region \citep{Sime1989,Komm1993torsional,Imada2020,Sharma2021,Edwards2022,Zhang2023} in a pursuit to explore the role of solar magnetic activity in driving the rotation of the solar atmosphere. The reported link in their respective findings suggests a cyclic behaviour in equatorial rotation and a differential nature in these layers, similar to the solar cycle but with a lag \citep{Sharma2021,Zhang2023}. But such a pursuit was too riddled with further contradictions as other studies \citep[e.g.,][]{Li2012,Bertello2020,Mishra2024} reported finding no such significant variation in the rotation rate of the solar chromosphere \citep{Mishra2024} and corona with the solar cycle. The contrasting findings from these studies have been explored through various theoretical and analytical perspectives. Among those perspectives, the potential connection between the solar interior and the solar atmosphere through magnetic fields has been consistently proposed in many studies to resolve some of these perplexing results \citep[e.g.][]{Weber1969,wang1989magneticflux,Badalyan2005,Bagashvili2017,FinelyBrun2023}.

Despite utilizing various methods and datasets, a comprehensive understanding of the global rotational profile of the solar atmosphere above the photosphere and how it varies across different layers remains elusive due to the diverse results obtained. In an attempt to address this gap, this study adopts a more focused approach by utilizing a single tracer-independent method, that is, the method of image correlation, to analyze the extensive dataset provided by the Atmospheric Imaging Assembly (AIA) of the solar Dynamic Observatory (SDO) from the period of 2010--2023. Additionally, we utilize the internal rotation rates derived using helioseismology from \cite{Antia1998,Antia2008}, photospheric rotation rate using sunspot \citep{Jha2021}, and chromospheric rotation rate using chromospheric plage \citep{Mishra2024}, to connect the global variation of the solar differential rotation profile from subsurface regime to the atmosphere.
In \autoref{sec:data}, we will discuss the specific details of the dataset used; \autoref{sec:Method} will discuss the modifications made to the method initially proposed by \cite{Mishra2024} for this study before moving on to \autoref{sec:results} and \autoref{sec:discussion} where results obtained from the analysis will be discussed. The final \autoref{sec:conclusion} will summarize the study and highlight its key conclusions.

\section{Data} \label{sec:data}

The Atmospheric Imaging Assembly \citep[AIA;][]{Lemen2012} on the Solar Dynamics Observatory \citep[SDO;][]{ChamberlinSDO2012,Pesnell2012} captures data across multiple extreme ultraviolet (EUV) and ultraviolet (UV) wavelengths. By employing narrow-band imaging in ten specific temperature-sensitive wavelength channels, such as Fe {\sc xviii} ($94$\,{\AA}), Fe {\sc viii, xxi} ($131$\,{\AA}), Fe {\sc ix} ($171$\,{\AA}), Fe {\sc xii, xxiv} ($193$\,{\AA}), Fe {\sc xiv} ($211$\,{\AA}), He {\sc ii} ($304$\,{\AA}), and Fe {\sc xvi} ($335$\,{\AA}), the AIA probes the solar atmosphere at different temperatures ranging from $\approx 10^4$ K to $10^7$ K. The AIA observes regions of solar atmosphere starting from the photosphere and above, extending through the chromosphere, transition region, and lower corona, with a pixel scale of $0.6${\arcsec}/pixel. In addition, one of the telescopes of the AIA observes in C {\sc iv} line near $1600$\,{\AA} and the nearby continuum at $1700$\,{\AA} as well as in the visible continuum at $4500$\,{\AA} \citep{Lemen2012}. The AIA has been providing data from May 2010 to the present, covering solar cycle 24 and ongoing cycle 25.

For this study, we primarily utilize data from seven different wavelength channels, namely \sixteenhundred, \threezerofour, \onethirtyone, \oneseventyone, \oneninetythree, \twooneone\, and \threethirtyfive\, from the period of 2010--05-13 to 2023--08--30 at a cadence of 6 hours. This was done to ensure that only features with a lifespan longer than 6 hours contributed to the analysis while still providing sufficient data for robust statistical analysis. The $94$\,{\AA} band was excluded from the analysis due to its poor signal-to-noise ratio in the low-temperature regime \citep{Aschwanden2013,Nuevo2015}. The initial dataset, classified as Level 1, is obtained from the Joint Science Operations Center\footnote{AIA data can be downloaded from \href{http://jsoc.stanford.edu/ajax/exportdata.html}{here}} \citep[JSOC;][]{jsoc1997} and converted to level 1.5  using Interactive Data Language (IDL) version of {\it aia\_prep.pro}, available under AIA/SolarSoft \citep{Solarsoft1998}. This step aligns the solar north with that of the image and applies the necessary adjustments required to make the plate scale consistent across all wavelength bands \citep{Lemen2012}. This Level 1.5 data from 2010--05-13 to 2023--08--30 was used to get the rotation profile of the Sun across all the aforementioned wavelengths.

\section{Methodology}\label{sec:Method}
\subsection{Pre-Processing}

\begin{figure}[htbp!]
    \includegraphics[width=\columnwidth]{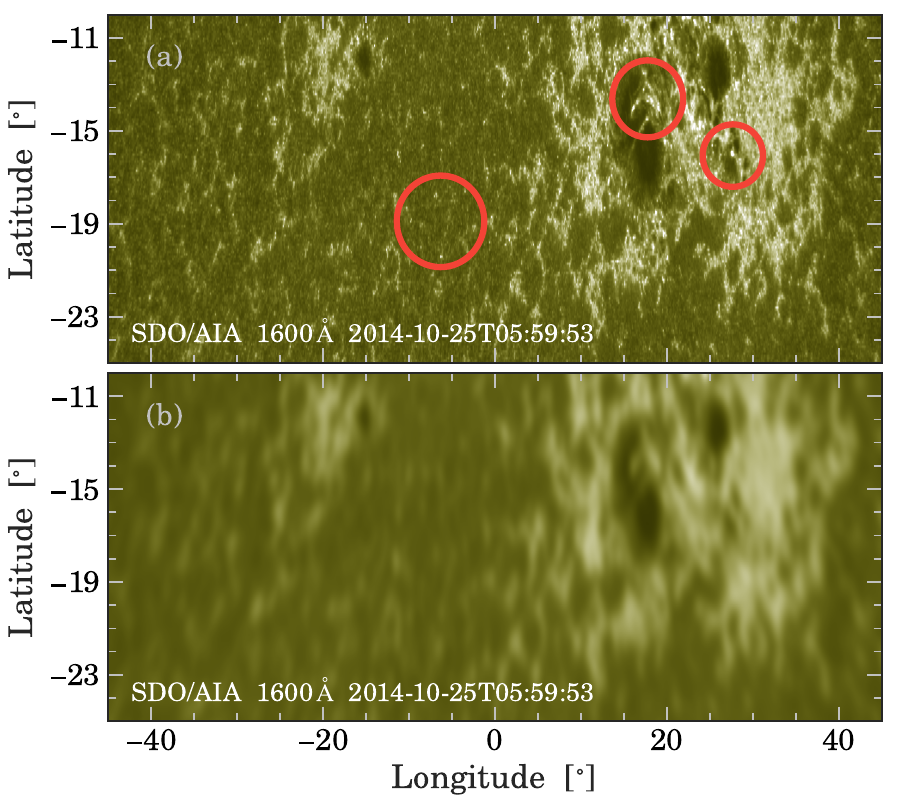}
    \caption{An example pair of images showing the effectiveness of Gaussian smoothing. (a) The level 1.5 data from \sixteenhundred\, with small scale brightenings, network bright points and fracture in plage regions. (b) The data after smoothing, as a result of convolution with a Gaussian kernel.}
    \label{fig:effectofsmoothing}
\end{figure}

The hotter solar atmosphere is ubiquitously populated with small-scale features that are often short-lived and can undergo significant changes within a very short span of time \citep{Solanki1993smallscale,bhatnagarbook2005}. Since the method of image correlation depends only on pixel-specific intensities in consecutive images, such rapid changes in the small-scale structures negatively affect
the correlation coefficient, thereby affecting our analyses. Therefore, the data was smoothed by convolving them with a Gaussian kernel to remove such small scale features to minimize the contribution from them. The size of the Gaussian kernel ($\sigma=5${\arcsec}) was chosen keeping in mind the angular size of the small-scale features \citep[$\Delta\theta \approx 1${\arcsec}- 5{\arcsec} e.g., chromospheric network and internetwork, quiet Sun concentrations;][]{Pozuelo2023}.
Additionally, this procedure serves as a step to minimize the random noise and help us to improve the signal-to-noise ratio of large-scale structures \citep{GaussSmooth2012}, see \autoref{fig:effectofsmoothing} for a representative example.

\subsection{Method of Image Correlation}
After applying a Gaussian smoothing filter, we utilized the image correlation technique similar to \cite{Mishra2024} to determine the rotation rate in different latitude bands. The image correlation method utilizes the two-dimensional (2D) cross-correlation technique to determine the offset between two images. This method has been previously suggested to focus on the rotation of the magnetic features as has been discussed in \cite{Snodgrass1983,Snodgrass1992} and \cite{stenflo1989}. The method is briefly outlined below, but for a detailed discussion, the reader is encouraged to refer to \cite{Mishra2024}.

\begin{figure}[!htbp]
\centering
\includegraphics[scale=0.6]{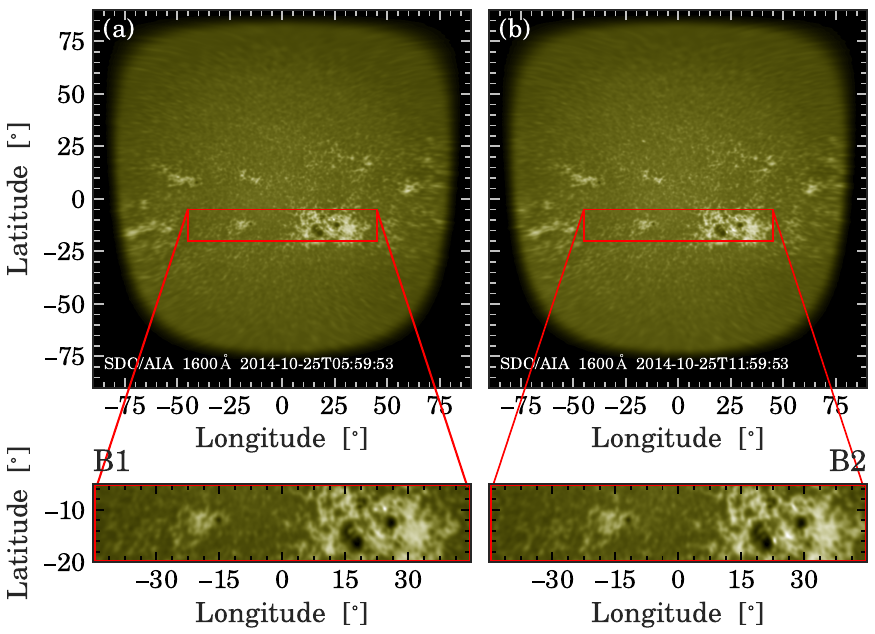}
    \caption{Panels (a) and (b) depict an example pair of images, temporally separated by 6 hours, after projection onto the heliographic grid. Red rectangular boxes in (a) and (b) represent the selected bands (B1 and B2) for cross-correlation, spanning $-20^{\circ}$ to $-5^{\circ}$) in latitude and $\pm 45^{\circ}$ in longitude in this example.}
    \label{fig:heliodemonstration}
\end{figure}

We project the full-disk AIA data to a heliographic grid of size 1800\,pixels\,$\times$\,1800\,pixels ($0.1^\circ$/pixel in latitude and longitude) using the near-point interpolation; see \autoref{fig:heliodemonstration}(a), similar to the process demonstrated in \cite{Mishra2024}. These projected images are then divided into overlapping bins of $15^{\circ}$, each separated by a $5^{\circ}$ stride e.g., $0^{\circ}-15^{\circ}$, $5^{\circ}-20^{\circ}$ etc. (see \autoref{fig:heliodemonstration}). The choice of $15^{\circ}$ bin is made to minimize the impact of any partially remaining extended features \citep{Weber1999,Meunier2003,Riha2007} and improve the cross-correlation coefficient. Furthermore, the overlapping bins are chosen to ensure a sufficient number of latitudinal bands were probed. These bins are selected over the span of $\pm 60\degr$ (in the case of \oneseventyone, \oneninetythree, and \twooneone), $\pm 55\degr$ (in the case of \sixteenhundred, \threezerofour) and $\pm 45\degr$ (in the case of \threethirtyfive\, and \onethirtyone) in latitude ($\theta$) and $\pm 45^{\circ}$ in longitude ($\phi$). These multiple latitudinal extents are selected to take into account the presence of most of the large-scale features across different wavelength channels e.g., active regions, large-scale coronal bright points (CBPs) within $\pm 45\degr$, plages within $\pm 55\degr$. Additionally, these limits also serve to reduce the projection effects at higher latitudes ($\theta>\pm 60\degr$) and near the limb \citep{Weber1999,Deforest2004}. The latitude of the bin is assigned as the centre of the selected bin, e.g., for $0^{\circ}-15^{\circ}$ it is $7.5^{\circ}$. Subsequently, two bins (say B1 and B2) of the same latitudinal extent from consecutive images (separated by 6\,hrs in time) are used to calculate the 2D cross-correlation function by shifting B2 with respect to B1 for the set of $\Delta \phi \in [ \phi_{0}-3^{\circ}, \phi_{0}+3^{\circ}]$ in longitude and $\Delta \theta \in [\theta_{0}-1^{\circ}, \theta_{0}+1^{\circ}]$ in latitude direction, where $\phi_{0}$ is the expected longitudinal shift estimated based
on the photospheric rotation rate \citep{Jha2021} and $\theta_{0}$ is taken as 0. Finally, the $\Delta \phi$ and $\Delta \theta$ are identified by maximizing the 2D cross-correlation function\footnote{The image cross-correlation was performed using \href{https://hesperia.gsfc.nasa.gov/ssw/gen/idl_libs/astron/image/correl_images.pro}{correl\_images.pro} and \href{https://hesperia.gsfc.nasa.gov/ssw/gen/idl_libs/astron/image/corrmat_analyze.pro}{corrmat\_analyze.pro} routines available in the Solar SoftWare library.}. Since this study is focused towards the measurement of differential rotation particularly, only the value of $\Delta \phi$ was used to calculate the value of $\Omega$ in that latitudinal bin.

\section{Results}\label{sec:results}

\subsection{Average rotational profile above the photosphere}\label{subsec:averagerotprof}

\begin{figure*}[!htbp]
    \centering
    \includegraphics[width=\textwidth]{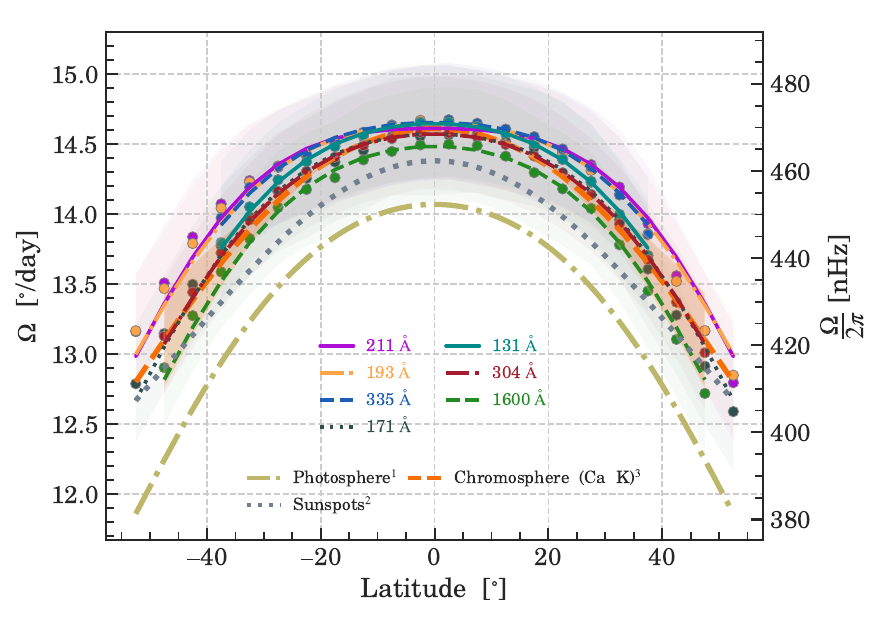}
    \caption{The average rotational profiles of all AIA channels starting from the chromosphere to corona, along with the results from $^{1}$\cite{Snodgrass1984}, $^{2}$\cite{Jha2021} and $^{3}$\cite{Mishra2024}.}
    \label{fig:allrotprof}
\end{figure*}

To obtain the average rotational profile of the hotter solar atmosphere above the solar photosphere, we calculated the average of $\Omega (\theta)$ for each latitudinal band weighted by corresponding cross-correlation coefficients (CC) in that latitude band. This step is performed after the elimination of cases with low values of CC, which may have arisen due to the absence or emergence of any large-scale feature in either of the consecutive images being analyzed. Low values of CC may also result from the presence of transient events (e.g., flares), which lead to intensity enhancements in any of the consecutive images being correlated. Cases where CC $< 0.65$ for \onethirtyone, \oneseventyone, \oneninetythree, \twooneone, \threethirtyfive\, and CC $<0.70$ for \threezerofour\, and \sixteenhundred\, are not included in analysis. These limits on the CC are imposed after finding out the value of CC for which the values of $A, B$ and $C$ do not vary significantly \citep[for detailed discussion on this approach see][]{Mishra2024}. The uncertainty (error) in $\Omega_{\theta}$ is calculated as the resultant of the least count error ($\sigma_{{\rm LCE}}$) and the standard statistical error ($\sigma_{{\rm SSE}}$) of the mean. However, $\sigma_{{\rm LCE}}$ remains dominant in the total error estimate\footnote{$\sigma_{{\rm LCE}}$ = $\frac{\Delta\phi}{\Delta t} = \frac{0.1^{\circ}}{0.25 {\rm days}} = 0.4^{\circ}$}, as $\sigma_{{\rm LCE}}$ is an order of magnitude greater than the $\sigma_{{\rm SSE}}$ (shaded region of respective colours for each wavelength band in \autoref{fig:allrotprof}). The values of mean $\Omega_{\theta}$ thus obtained for each latitudinal bin are then fitted with \autoref{diffequation} (where $\theta=\theta_{mid}$ is the centre of the latitude band) using the least square fit method, to obtain the best-fit parameters ($A$, $B$, $C$) and their associated uncertainties ($\Delta A$, $\Delta B$, $\Delta C$). These steps are repeated for each AIA wavelength channel, and the differential rotation parameters obtained are tabulated in \autoref{tab:All wavelength results} for the same. 

\begin{deluxetable*}{ccccccc}
    \tabletypesize{\small}
\tablewidth{0pt}
\tablecaption{The values of differential rotation parameters for different wavelength channels}
 \tablehead{\colhead{Wavelength (Primary ion)} &\colhead{Log$_{10}\,T$*} & \colhead{Height $\pm$ error*} & \colhead{$A \pm\Delta A$ } & \colhead{$B \pm\Delta B$}& \colhead{$C \pm\Delta C$}\\
\colhead{ (\AA)} & \colhead{}  & \colhead{(z, km)}  & \colhead{($^{\circ}\rm/day$)} & \colhead{($^{\circ}\rm/day$)} & \colhead{($^{\circ}\rm/day$)} 
 }
\startdata 
         304 (He {\sc ii}) & 4.7 & 2820 $\pm$ 400 & $14.574 \pm 0.012$ & $-1.518 \pm 0.12$ & $-2.287 \pm 0.223$ \\
        1600 (C {\sc iv}) & 5.0 &  430 $\pm$ 185 & $14.485 \pm 0.024$ & $-1.612 \pm 0.243$ & $-2.677 \pm 0.452$ \\
         131 (Fe {\sc viii}) & 5.6 & --- & $14.649 \pm 0.014$ & $-1.334 \pm 0.200$ & $-2.999 \pm 0.519$ \\
         171 (Fe {\sc ix}) & 5.93 & 5100 $\pm$ 1900 & $14.574 \pm 0.032$ & $-1.356 \pm 0.286$ & $-2.654 \pm 0.461$ \\
         193 (Fe {\sc xii}) & 6.176 & 6700 $\pm$ 2000 & $14.645 \pm 0.039$ & $-0.916 \pm 0.344$ & $-2.701 \pm 0.557$ \\
         211 (Fe {\sc xiv}) & 6.272 & 6100 $\pm$ 1900 & $14.613 \pm 0.042$ & $-0.504 \pm 0.372$ & $-3.314 \pm 0.601$ \\
         335 (Fe {\sc xvi}) & 6.393 & 15200 $\pm$ 2300 & $14.656 \pm 0.015$ & $-0.958 \pm 0.224$ & $-2.750 \pm 0.604$ \\
\enddata
\tablecomments{*The logarithmic temperatures and heights used to represent all wavelength channels are obtained from the studies of \cite{Simon1972,Simon1974,Fossum2005,Kwon2010,Howe2012,Lemen2012,Nuevo2015}. A detailed discussion is available in \autoref{appendix:aiaheightest}.}
\label{tab:All wavelength results}
\end{deluxetable*}

Our first interpretation from \autoref{fig:allrotprof} is that the rotation profile in the solar atmosphere, from the upper photosphere to the corona, exhibits a similar rotational profile across all wavelengths. However, the rotational profile is relatively flatter i.e less differential whereas the absolute rotation rate is higher compared to the photosphere,
as observed using Dopplergrams \citep[olive green dash-dotted line in \autoref{fig:allrotprof};][]{Snodgrass1984,Ulrich1996} and sunspots as tracers \citep[dark grey dotted line in \autoref{fig:allrotprof};][]{Jha2021}. The rotational profile corresponding to chromospheric temperatures (\threezerofour) aligns well with the findings of \cite{Mishra2024}, adding credibility to our results. Additionally, our results are consistent with the \cite{Chandra2010, Morgan2011, Sharma2020, Sharma2021, Edwards2022}, suggesting that the corona rotates faster and less differentially (see \autoref{tab:All wavelength results}). 


\subsection{Variation of rotational parameters with height and temperature}\label{subsec:height}

In order to investigate the variations in solar differential rotation from the photosphere to the corona, as is indicated in studies like \cite{Vats2001,Altrock2003,Sharma2020,Imada2020}, it is necessary to get the corresponding height of all AIA channels. Several 1D models of the solar atmosphere through the photosphere to the transition region have been proposed throughout the years \citep{Vernazza1981,Fontenla1993}. However, these models provide features that seldom agree with the observed profiles due to several factors \citep{Avrett2008}. Hence, we obtain the approximate representative heights above the photosphere to represent the parts of the solar atmosphere visible in the wavelength channels used in this study (see \autoref{tab:All wavelength results}), keeping in mind the temperature sensitivity of the same \citep{Simon1972,Simon1974,Fossum2005,Kwon2010,Howe2012}. Unfortunately, to the best of our knowledge, there was no singular height that could be ascribed to the AIA \onethirtyone\, 
Consequently,  we have not included the rotation parameters measured using data from this channel in this part of the analysis. Furthermore, in order to make a fair comparison of rotational parameters across all wavelength bands, we also study the variations in rotational parameters with the temperature ($T$) corresponding to each wavelength.

\begin{figure*}
    \centering
    \gridline{\fig{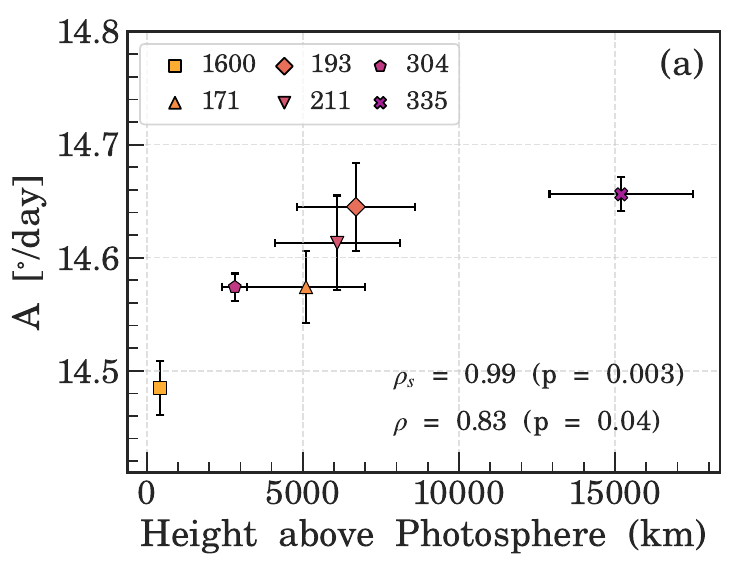}{0.5\textwidth}{}\fig{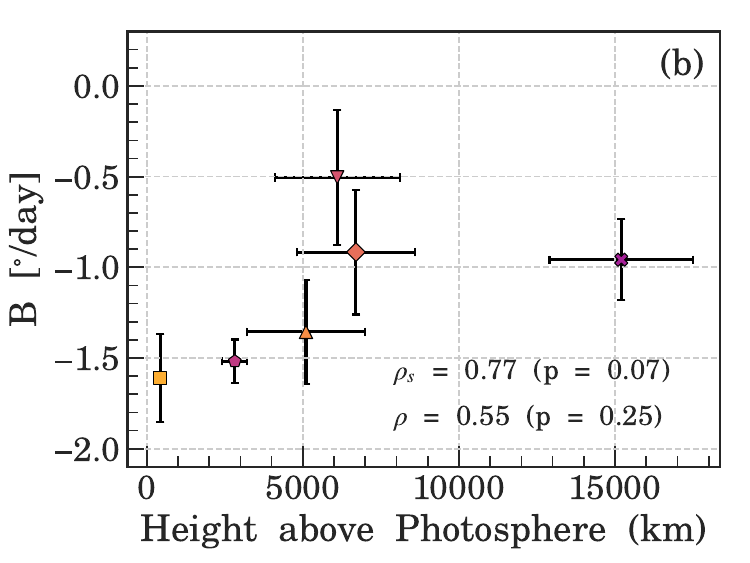}{0.5\textwidth}{}}
    \vspace{-0.5cm}
    \gridline{\fig{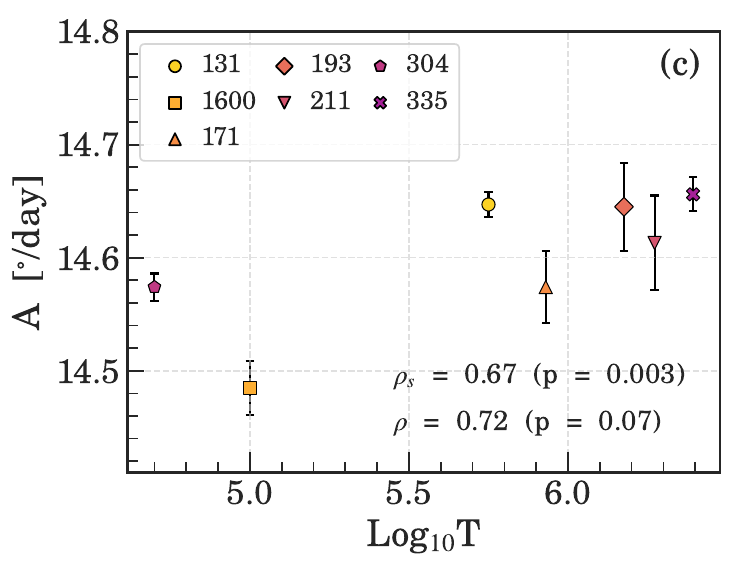}{0.5\textwidth}{}\fig{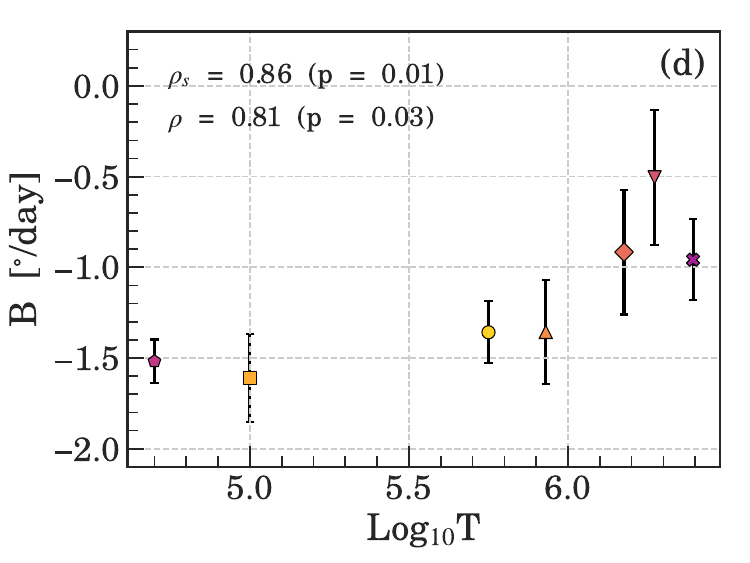}{0.5\textwidth}{}}
    \vspace{-0.8cm}
    \caption{The trend in (a) equatorial rotation rate (A), and (b) differential gradient (B) with increasing height above the photosphere. All error bars in the $y-$axes correspond to the uncertainty pertained in determining the parameters $A$ and $B$, whereas the errorbars along the $x-$axis correspond to the errors as determined by the original studies as listed in \autoref{tab:All wavelength results}. The variation in (c) Equatorial rotation rates (d) Differential gradients of the solar atmosphere as obtained with temperature.}
    \label{fig:height}
\end{figure*}

In \autoref{fig:height}, we plot the rotational parameters, $A$ (\autoref{fig:height}a) and $B$ (\autoref{fig:height}b), against the height ($z$) above the photosphere and $T~(\log_{10}T)$, whereas $B$ with $z$ and $T$ in \autoref{fig:height}c, \autoref{fig:height}d, respectively. Finally, to assess the extent of the relationship between the said parameters, we calculated both Spearman ($\rho_s$) and Pearson ($\rho$) correlation coefficients (CC) between rotational parameters and the $\log_{10}T$.
These positive values of CC, between ($A, z$) and ($B, z$), are indicative of an increase in equatorial rotation as well as the decrease in latitudinal gradient (flatter profile) with height in the solar atmosphere, which has been previously speculated by \cite{Parker1982,Japaridze1992,wang1988,Vats2001,Altrock2003,Sharma2020}. Additionally, the relationship between $A$ and $\log_{10}T$ seems to exhibit similar behaviour, although with lower CC. Here we would like to emphasize the scarcity of data in the temperature range from approximately $\log_{10}T=5.0$ (represented by \sixteenhundred) till $\log_{10}T\approx5.9$ (represented by \oneseventyone). This absence of information could potentially have a significant impact on the determination of the correlation in the present scenario. Conversely, the $B$ shows an upward trend in connection with $\log_{10}T$. Based on \autoref{fig:height}, we note that although the rotation parameters show positive CC, the nature of the increase is different in these two cases (temperature and height).

\subsection{Is there a connection with the solar interior?}\label{subesc:helioseism}

It is imperative to acknowledge from \autoref{fig:allrotprof} that the rotation rate of the solar atmosphere (for all AIA channels) is faster than that of the rotation rate measured using photospheric magnetic features like sunspots. Interestingly, the rotation rate derived using magnetic features, which are believed to be anchored deeper in the photospheric surface, is greater than 
the rotation rate obtained based on Doppler measurement, which samples the higher photospheric plasma \citep{Komm1993,Komm1993torsional,Xiang2014,Xu2016}.
Such results motivate us to consider the potential connection between the faster-rotating solar interior to the faster-rotating solar atmosphere measured in this study. In order to explore such possibilities, we need to obtain the profile of the Sun's rotation from the subsurface regime to its outer layers, observing how rotational characteristics evolve from the interior to the atmosphere of the Sun.

\begin{figure}[!htbp]
\epsscale{1.20}
    \plotone{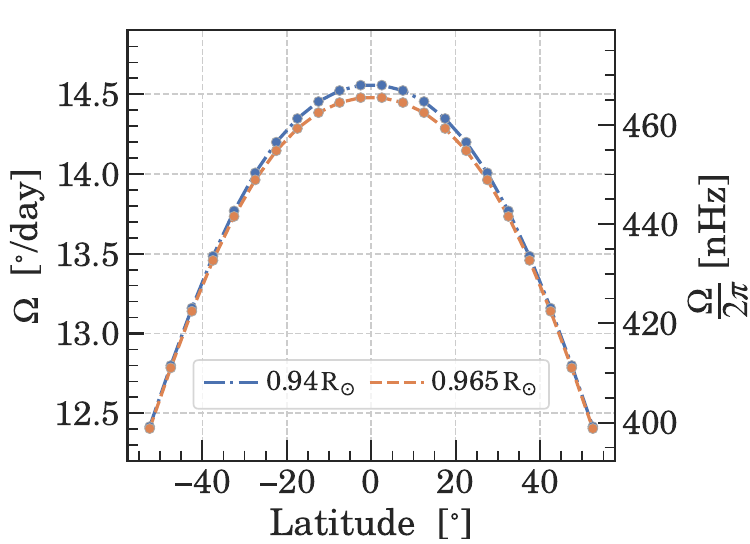}
    \caption{A representative plot of the rotational profiles obtained from helioseismological data assuming a symmetric distribution of rotation rates in both hemispheres.}
    \label{fig:rotcurveshelio}
\end{figure}

We have already obtained the rotation rate of the solar atmosphere whereas for internal rotation, we use the helioseismic measurement of solar rotation, obtained using the methodology outlined in \cite{Antia1998,Antia2008}. The helioseismic data we use is the temporally averaged values of $\Omega (r,\theta)$ for $r\in [0.7\,{\rm R}_{\odot},1.0\,{\rm R}_{\odot}]$ in steps of $0.005\,{\rm R}_{\odot}$ and $\theta\in[0^{\circ}-88^{\circ}]$ in steps of $2^{\circ}$. To obtain the rotation parameters i.e., $A$, $B$ and $C$ for a given depth $r$ we fit $\Omega (r,\theta)$ with \autoref{diffequation} for latitudes spanning $\pm 60^{\circ}$ (as most of the solar magnetic features considered in this study are limited within this latitude). This calculation is only performed for all $r\in(0.93,1.0)\,{\rm R}_{\odot}$, as we are assuming the possibility of sub-photospheric influence on solar atmospheric rotation. A representative example of rotation profile for $r=0.94\,{\rm R}_{\odot}$ (deeper)  and $r=0.965\,{\rm R}_{\odot}$ (near the surface) is shown in \autoref{fig:rotcurveshelio}.

\begin{figure*}
\centering
    \includegraphics[scale=0.6]{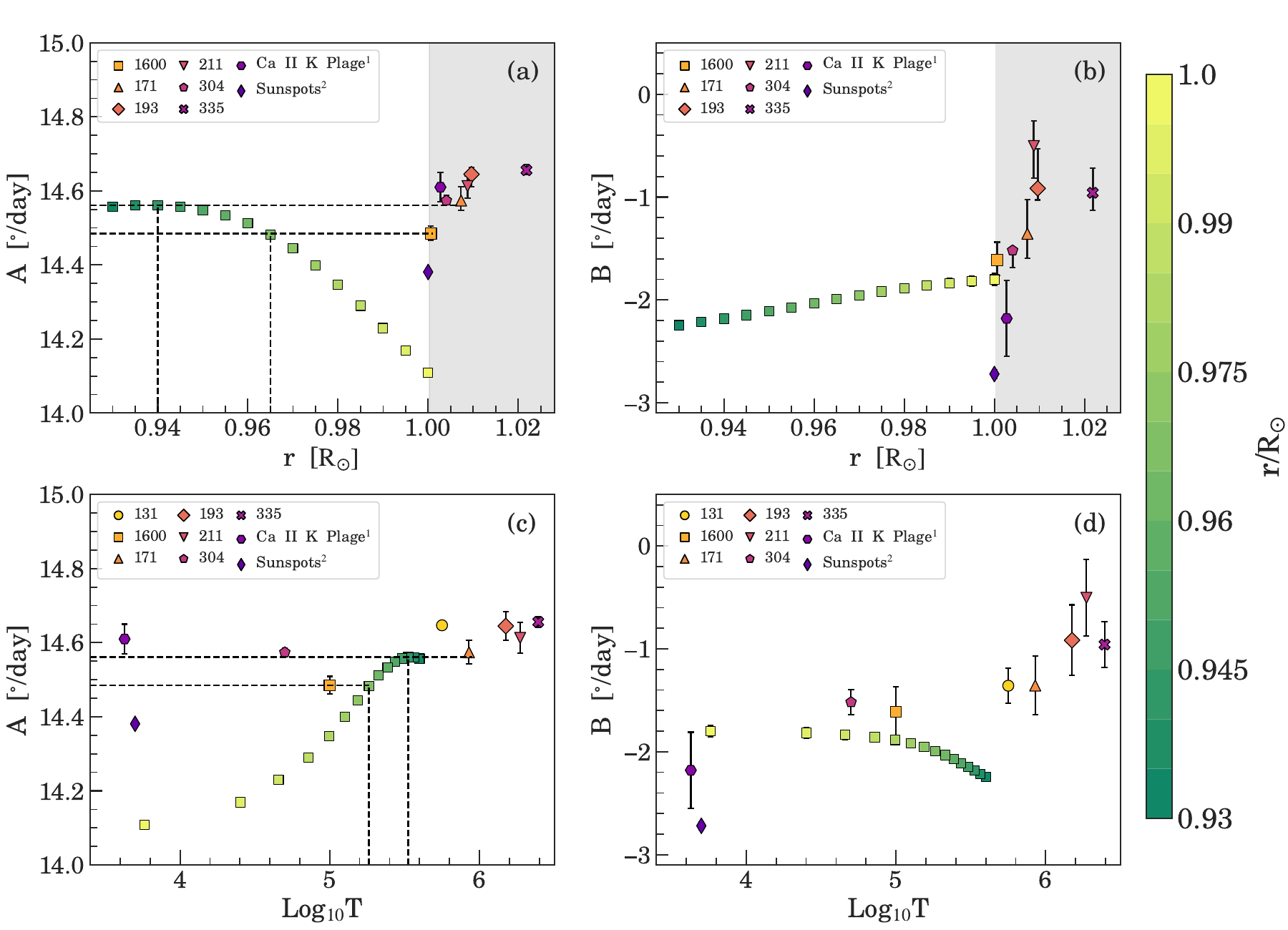}
    \caption{The variation in rotational parameters Equatorial Rotation rate ($A$) and Latitudinal gradient ($B$) from the interior to the atmosphere (highlighted in grey) of the Sun as a function of radius (top panel; (a) and (b)) and logarithmic temperature (bottom panel; (c) and (d)), as obtained from this study, $^{1}$\cite{Mishra2024}, $^{2}$ \cite{Jha2021} and helioseismology. The height representative of Ca {\sc ii} K plages is obtained from \cite{Stix1976} while the logarithmic temperature is obtained from \cite{Beebe1971Formation}}
    \label{fig:all_variation_with_helioseismology}
\end{figure*}

To examine the evolution in rotational parameters from sub-photspheric regime to the atmospheric values obtained in this study, we compare our results with the rotation rate inferred from helioseismology with respect to depth as well as the corresponding temperature (see \autoref{fig:all_variation_with_helioseismology}), derived from Solar S-Model \citep{christensen1996current}. On comparing our results in \autoref{fig:all_variation_with_helioseismology}a, interestingly, we note that the rotation parameter $A$ for solar atmosphere as obtained from the \threezerofour\, and \oneseventyone\, coincide with the rotation parameter $A$ obtained at a depth of $r \approx 0.94\,{\rm R}_{\odot}$ as well as the $A$ for \sixteenhundred\, coincides with the $A$ for $r \approx 0.965\,{\rm R}_{\odot}$. Additionally, we also note that the $A$ at $r = 0.94\,{\rm R}_{\odot}$ also show a good match with the $A$ obtained in the case of \twooneone, if we consider $3\sigma$ uncertainty for $A$. At this juncture, it is imperative to emphasize that the \twooneone\, channel receives a contribution from cooler components too, with the temperature near to the one the \oneseventyone\, channel is sensitive to. This further highlights the complexity of considering the solar atmosphere to be distinctly stratified, with the contribution from each layer being unique and independent. We acknowledge the importance of considering the potential contributions from different heights in the same channel when determining equatorial rotation rates, as demonstrated by the case of \twooneone.


\begin{figure*}[!htbp]
    \centering
    \includegraphics[scale=0.55]{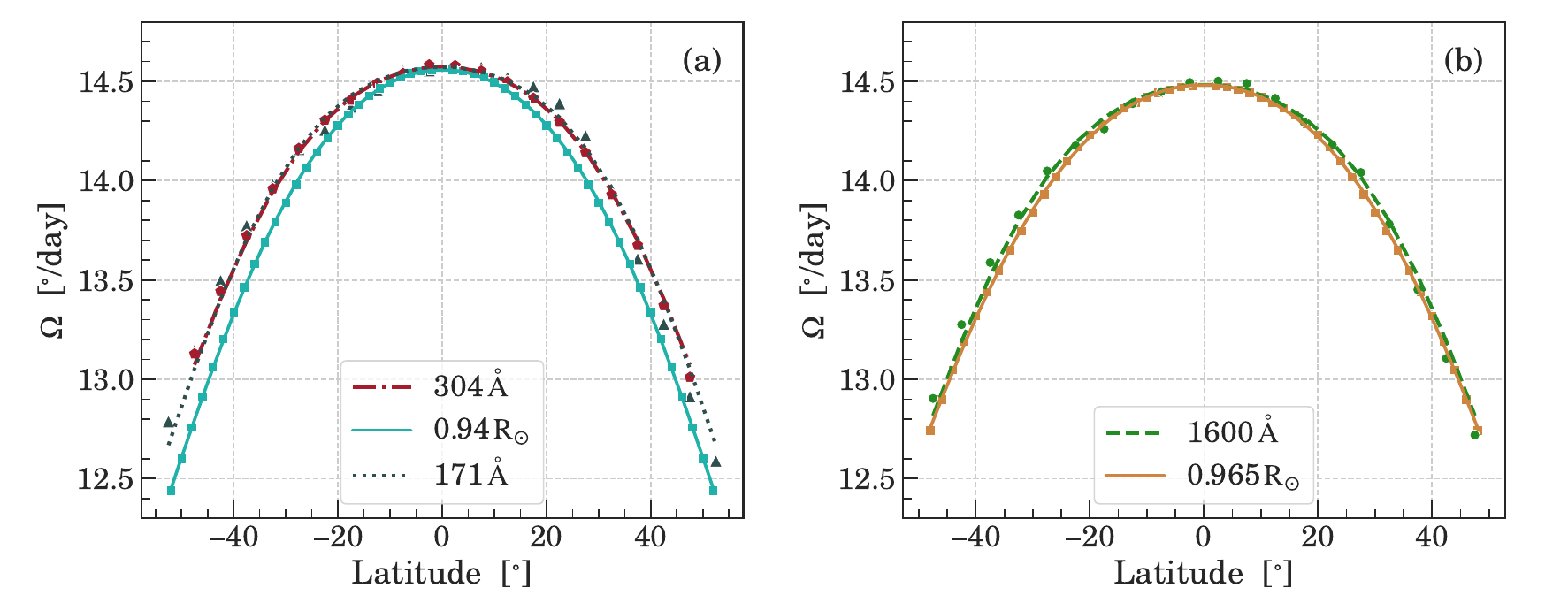}
    \caption{Comparison of the rotational profile at the depth of (a) $0.94\,$R$_{\odot}$, with that obtained for \oneseventyone\, and \threezerofour; and (b) \sixteenhundred\, with $0.965\,$R$_{\odot}$ at all latitudes.}
    \label{fig:171and1600heliocorrel}
\end{figure*}

Considering the complexity associated with the determination of unique height, in \autoref{fig:all_variation_with_helioseismology}c-d, we plot $A$ and $B$ obtained for atmosphere as well as interior as a function of $T~(\log_{10}T)$ instead of the $z$. Interestingly, we find the exact same match with the internal rotation for these wavelengths. However, for $B$ we do not find any such clear connection between the interior and atmosphere of the Sun (see \autoref{fig:all_variation_with_helioseismology}b and d).  On plotting the respective rotational profiles for the depths of $0.94\,$R$_{\odot}$ and $0.965\,$R$_{\odot}$ and channels \oneseventyone, \threezerofour\, and \sixteenhundred, we find a good overlap of the profiles for $0.965\,$R$_{\odot}$ and \sixteenhundred\, at all latitudes, while for  $0.94\,$R$_{\odot}$, \oneseventyone, \threezerofour, the overlap is more apparent at the equatorial regime (see \autoref{fig:171and1600heliocorrel}).

Such an overlap in equatorial rotation rates had also been previously discussed in \cite{Badalyan2005, Mancuso2020}, who had used Coronal Green Line Brightness (CGLB) data and ultraviolet (UV) spectral line observations, respectively, to obtain the rotational profile of the solar corona. Furthermore, \textcolor{blue}{\bf{\cite{Ruzdjak2004}}} had also previously suggested the anchoring of sunspots at $0.93\,$R$_{\odot}$ on a similar comparison with helioseismology results.



\subsection{Variation of rotational parameters with solar activity}\label{subsec:cyclicvar}
\begin{figure*}
\epsscale{1.15}
    \plottwo{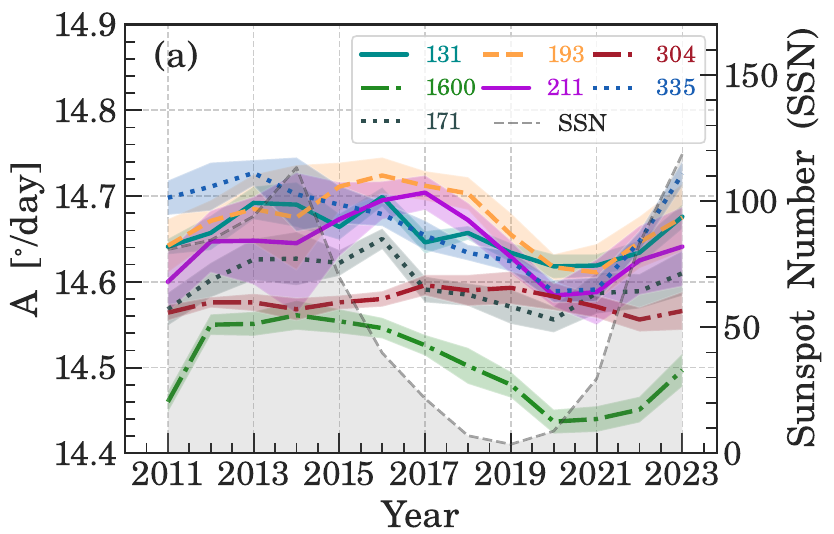}{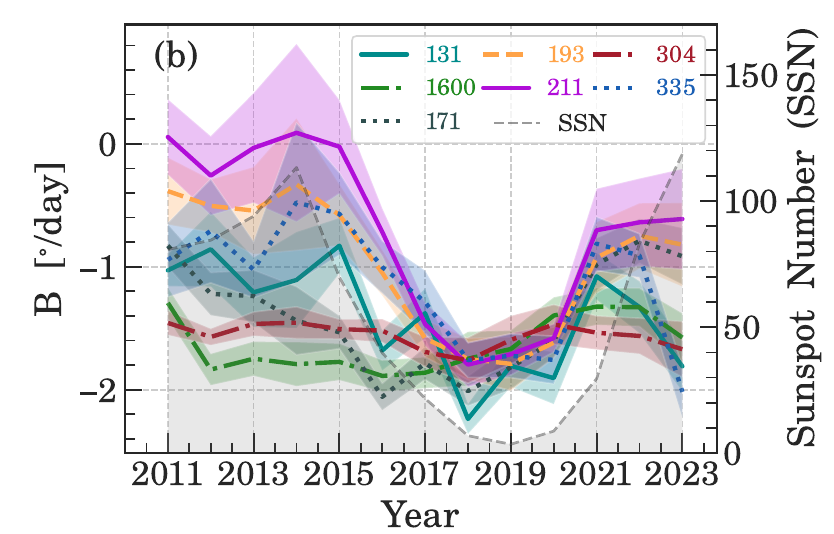}
    \caption{The variation in rotational constants with the progression of cycles 24 and 25 compared with the sunspot number. The shaded area in each colour represents the uncertainty in determining the respective parameters.}
    \label{fig:cyclevar}
\end{figure*}

Another topic that has persistently generated significant interest and debate is the impact of solar activity on the rotation rate of the Sun. Although the limited data span makes such a study challenging, we explore whether the rotational parameters of the solar atmosphere vary with solar activity, i.e., with the different phases of the solar cycle. To achieve this, we obtained the differential rotation parameters ($A$, $B$ and $C$) for each year using a similar approach as discussed in Section~\ref{sec:results}. These parameters are then plotted in \autoref{fig:cyclevar} as a function of time along with the yearly averaged sunspot number (SSN)\footnote{\href{https://www.sidc.be/SILSO/datafiles}{https://www.sidc.be/SILSO/datafiles}}, which is a marker for solar activity.

In \autoref{fig:cyclevar}, we note an apparent cyclic behaviour with time in the rotational parameters, both $A$ and $B$. To quantify this behaviour, we calculate the Spearman rank correlation ($\rho_s$) of the parameters obtained for each channel with the SSN (see \autoref{tab:cyclicvarcorr}). We note that the results from the channel \sixteenhundred\, as well as the channels with sensitivity to temperatures native to coronal heights (\oneseventyone, \oneninetythree, \twooneone, \onethirtyone\, and \threethirtyfive) show a significant positive correlation in equatorial rotation rate, $A$, with solar cycle phase (p$ <0.05$). Such a correlation of the differential rotation of different parts of the solar corona with the solar activity cycle has been reported in many studies \citep{Vats1998cyclevar,Li2012,Javariahcycle2013,Jurdana2011,Xie2017coronalrot,Imada2020}. In contrast, the chromospheric channel (\threezerofour) shows a very low and negative value of CC, thereby barely indicating any variation with solar activity. This behaviour of the chromospheric rotational profile is consistent with the results for chromospheric rotation as obtained in \cite{Mishra2024} and \cite{Jha2021} for sunspot, who found no significant change in equatorial rotation rate with solar activity. However, on studying cyclic variation of differential rotation parameters using sunspot data from various databases, \cite{Ruzdjak2017} found that the equatorial rotation rate does reach its maximum just before solar activity minimum, which can be noticed from a careful comparison of the general trend of $A$ with solar activity.

For parameter $B$, which represents the differential nature of the rotation, we find no such significantly high positive correlation (p $<0.05$) in any channel other than \twooneone. \cite{Jurdana2011} obtained a similar lack of correlation for parameter $B$; however, they attributed this lack of correlation to more pronounced errors in their data at higher latitudes.

\begin{deluxetable}{ccccccc}
\tabletypesize{\small}
\tablewidth{0pt}
\tablecaption{Spearman rank correlation coefficients of the rotational parameters and yearly averaged Sunspot numbers (SSN).\label{tab:cyclicvarcorr}}
\tablehead{\colhead{Wavelengths} & \multicolumn{2}{c}{A} & \nocolhead{} & \multicolumn{2}{c}{B}\\
\colhead{ (\AA)} & \colhead{$\rho_{\rm s}$}  & \colhead{p-value} & \nocolhead{} & \colhead{$\rho_{\rm s}$} & \colhead{p-value}
}
\startdata
           304 & -0.309 & 0.304 & & 0.269 & 0.374 \\
          1600 & 0.863 & 0.001 & & -0.52 & 0.069 \\
          131 & 0.813 & 0.001 &  & 0.489 & 0.09 \\ 
           171 & 0.863 & 0.001 & & 0.148 & 0.629 \\
           193 & 0.583 & 0.036 &  & 0.67 & 0.012 \\
             211 & 0.571 & 0.041 &  & 0.725 & 0.005 \\
            335 & 0.83 & 0.0005 & & 0.462 & 0.112 \\
\enddata
\end{deluxetable}

\section{Discussion}\label{sec:discussion}

The rotational profiles of the upper solar atmosphere in \autoref{fig:allrotprof}, as seen in different wavelength regimes, suggest that the solar atmosphere, as modulated by magnetic large-scale features like plages, CBPs, filaments, coronal loops, etc., rotates $2.95\%-4.18\%$ and $0.73\%-1.92\%$ faster (at the equator) and less differentially compared to the photospheric rotational profile obtained from dopplergrams and sunspot data, respectively. However, these results are obtained based on the method of image correlation, which is sensitive to the intensity contrast of multiple magnetic features in the hotter solar atmosphere. While this method does not distinguish between the rotation of individual features, this method has the advantage of improving the statistics of the analysis by taking into account all the features distinguishable by intensity. The hotter solar atmosphere is also an optically thin region, and therefore, a measurement of shift in features could also be affected by the line-of-sight effect, leading to an apparent measurement of faster rotation. Therefore, there is a possibility that these results have an effect of the apparent line-of-sight (LoS) effect arising because of the extended structures like coronal loops. To test this hypothesis, we created a toy model of an extended structure mimicking a coronal loop to examine the extent of such an effect (see \autoref{appendix:LoSeffect}). The results obtained based on this experiment have confirmed that the difference between the photospheric rotation rates and that beyond the photosphere can not only be the outcome of the line-of-sight effect.  However, it might have a small effect on it, which we have quantified in \autoref{appendix:LoSeffect}. 

Once we have eliminated this prospect, it is important to acknowledge that the solar atmosphere is not uniformly stratified and is multi-thermal; consequently, a filter sensitive to a specific temperature may receive contributions from various heights. However, this study relies on the well-established understanding that certain global temperature ranges (e.g., $\approx 10^{6}$ K) are limited to the higher layers of the solar atmosphere (e.g., the solar corona). Although these temperatures may be instantaneously achieved locally in the lower layers of the solar atmosphere during transient events (e.g., flares), we assume they do not represent the long-term global characteristics of the solar atmosphere, which is the primary focus of this study. As a result, the outcomes derived in this study remain statistically unaffected by these events.

The observed increasing trend in the solar differential rotation with height is a very debatable topic, and the proper explanation for such behaviour is still incomplete. However, a theoretical perspective was proposed by \cite{Weber1969}, discussing the role of magnetic field line configuration on atmospheric rotation leading to an increasing rotation rate with height which is in agreement with our measurement. According to \cite{Weber1969}, the interplay between magnetic field torque and the velocity plasma flowing outward ensures that the rotation rate increases with an increase in radial distance to keep the total angular momentum conserved. Additionally, the role of the magnetic field in providing the angular momentum required for faster rotation of the solar atmosphere beyond the photosphere has also been suggested in many studies \citep{Komm1993,Badalyan2005,Kwon2010,Badalyan2010,Badalyan2018,Li2019,Imada2020,Edwards2022}

In Section~\ref{subesc:helioseism}, we find an excellent match between the equatorial rotational ($A$) rate at the depths of $r = 0.94\,{\rm R}_{\odot}$ and $r = 0.965\,{\rm R}_{\odot}$ as inferred in helioseismic observations, with that obtained for the channels \threezerofour, \oneseventyone\, and \sixteenhundred.  While this alignment may seem coincidental, the potential physical connection between them cannot be completely dismissed owing to Ferraro's law of isorotation \citep{Ferraro1937}, which hints toward such a possibility. According to this law, strong magnetic fields frozen in plasma tend to transport the angular momentum at their footpoints throughout their extent; thereby leading to a comparatively rigid nature of rotation in low $\beta$ plasma and we suspect that this could be the region behind the observed behaviour of solar rotation. The idea that the footpoints of the loops visible in \oneseventyone\, may have their root in the lower layers has been also suggested in works like that of \cite{Kwon2010}, while several works have also hinted at the possibility of sub-photospheric rooting of coronal magnetic features \textcolor{blue}{\bf{\citep{Zaatri2009,Bagashvili2017,Edwards2022,Kutsenko2022}}}. We emphasize here that, these arguments should be taken with a grain of salt and it needs a better and thorough study to confirm such possibility.

We have also observed a positive correlation between the changes in the rotation of the solar atmosphere and the phase of the solar cycle, as evidenced by the correlation with the yearly averaged sunspot number, more prominent in the rotational parameter $A$. Such a result could indicate a relationship between the solar atmospheric rotation and the presence of different magnetic structures (e.g., plages, coronal loops, CBPs etc.) during different phases of solar activity. Another possibility hints at the probable existence of a phenomenon called torsional oscillation, which has extensively been discussed and documented in various studies \citep[e.g.,][]{Komm1993torsional,Imada2020}. Notably, the possibility of such a phenomenon is prominently observed in the layers at coronal temperatures (\oneseventyone, \oneninetythree, \onethirtyone, \twooneone), while no such variation is observed in the chromospheric counterpart (\threezerofour) in agreement with the findings of \cite{Mishra2024} for the chromosphere and \citep{Jha2021} for sunspots. Additionally, the variation of the coronal rotational profile has been proposed to be affected by magnetic flux concentrations \citep{Komm1993torsional,Weber1999,Altrock2003,Imada2020}, which is positively correlated with solar activity.

Additionally from \autoref{fig:cyclevar}, it is also apparent that the parameter $A$ reaches its maximum just before the minimum, while the parameter $|B|$ has a greater value at cycle minimum for \twooneone, suggesting a more differential rotation at cycle minimum. This is similar to the results obtained by \cite{Ruzdjak2017} and can be suggestive of what is known as the braking effect exerted by non-axisymmetric magnetic fields \citep{Brun2004}. Theoretical efforts have been made to explain such a variation with cycle activity, results from which have highlighted the role of the strength of magnetic fields in the transport of angular momentum towards the equator \citep{Brun2004,BrunMiesch2004,Lanza2006,Lanza2007,Brun2009}.

Except \twooneone, no statistically significant correlation of the parameter $B$ with the yearly averaged sunspot number is apparent for most pass bands representing the hotter solar atmosphere. A possible connection can be made to the cross-talk between the parameters $B$ and $C$, which amplifies the noise-related uncertainties and obscures their actual time variation \citep{Snodgrass1984}. This cross-talk is also the reason why the individual variation in the parameter $C$ is not individually examined in the study.

\section{Summary \& Conclusion}\label{sec:conclusion}
In this study, we analyzed 13 years of SDO/AIA data to understand the solar atmospheric rotational profile, its variation at different layers of the solar atmosphere, and with parameters like temperature and solar activity. The primary conclusion we arrive at in this study is that the solar atmosphere, till lower coronal heights, rotates faster and less differentially compared to the photospheric rotation rates obtained from Dopplergrams and sunspot data.

The study also utilised data from helioseismology at different depths to understand the variation of the rotational profile from the interior to the atmosphere and subsequently found a significant correlation between the rotational rate at certain sub-photospheric depths ($0.94\,{\rm R}_{\odot}$,  $0.965\,{\rm R}_{\odot}$) and that obtained for the channels sensitive to certain temperatures of the solar atmosphere (\oneseventyone, \threezerofour\, and \sixteenhundred\, respectively).

While the current study has reinforced that the hotter solar atmosphere indeed does rotate faster and less differentially than the photosphere, numerous unanswered questions remain. Despite the few possibilities explored in this study, the physical understanding behind the observed increase in rotation rate and decrease in differential nature and their generalized trends with height above the photosphere, logarithmic temperature, and solar activity remains unclear. It is important to note that this study does not aim to provide detailed information about the rotational profiles at each specific temperature and height within each layer of the solar atmosphere but rather to provide an overview of the general trend in the rotation of the solar atmosphere from the photosphere to the chromosphere, transition region, and corona.

The findings of this study, if revisited with a larger dataset encompassing multiple cycles and a method capable of distinguishing between thermally distinct features at their exact height of formation, might have significant implications in our understanding of the overall behaviour of the Sun's differential rotation and its complex relationship with the solar magnetic field. A future study could focus on developing a method to isolate the high- and low-temperature components in images from each channel as well as isolating the specific height associated with them to map the variation in the rotational profile of the exact same feature at different heights of the solar atmosphere, which is crucial for a more thorough analysis. Further validation for the trends suggested in this study can also be provided through the use of a dataset that spans multiple solar activity cycles as well as through the use of orthogonalized fit functions. This will help mitigate any potential biases that may have arisen from using a dataset that spans fewer cycles.

\section{acknowledgments}
We thank the anonymous referee for carefully reviewing this study and providing valuable feedback on improving the manuscript further. S.R. is supported by funding from the Department of Science and Technology (DST), Government of India, through the Aryabhatta Research Institute of Observational Sciences (ARIES). The computational resources utilized in this study were provided by ARIES. The funding support for DKM's research is from the Council of Scientific \& Industrial Research (CSIR), India, under file no.09/0948(11923)/2022-EMR-I. TVD was supported by the C1 grant TRACE space of Internal Funds KU Leuven, and a Senior Research Project (G088021N) of the FWO Vlaanderen. Furthermore, TVD received financial support from the Flemish Government under the long-term structural Methusalem funding program, project SOUL: Stellar evolution in full glory, grant METH/24/012 at KU Leuven. The research that led to these results was subsidised by the Belgian Federal Science Policy Office through the contract B2/223/P1/CLOSE-UP. It is also part of the DynaSun project and has thus received funding under the Horizon Europe programme of the European Union under grant agreement (no. 101131534). Views and opinions expressed are however those of the author(s) only and do not necessarily reflect those of the European Union and therefore the European Union cannot be held responsible for them. The authors extend their gratitude to NASA/SDO, the SDO/AIA science team, and the Joint Science Operations Centre (JSOC) for the AIA data used in this study. We would also like to thank H.M. Antia for the helioseismology data and the Solar Influences Data Analysis Centre (\href{SIDC}{https://www.sidc.be/}) for the Sunspot data that was utilized in this study. TVD is grateful for the hospitality of DB and VP during his visit at ARIES in spring 2023. S.R., B.K.J. and D.K.M. also extend their gratitude to the Indian Network for Dynamical and Unified Solar Physicists (\href{https://sites.google.com/view/indus-solphys/home}{INDUS}) for overall support. This study has also utilized the resources of NASA Astrophysics Data System (\href{https://ui.adsabs.harvard.edu/}{ADS}) and Semantic Scholar (\url{https://www.semanticscholar.org/}) bibliographic services. 

\appendix
\section{Exploring the effect of Line of Sight (LoS) projection of extended structures on the result}\label{appendix:LoSeffect}

\begin{figure}[!h]
\epsscale{0.95}
\plotone{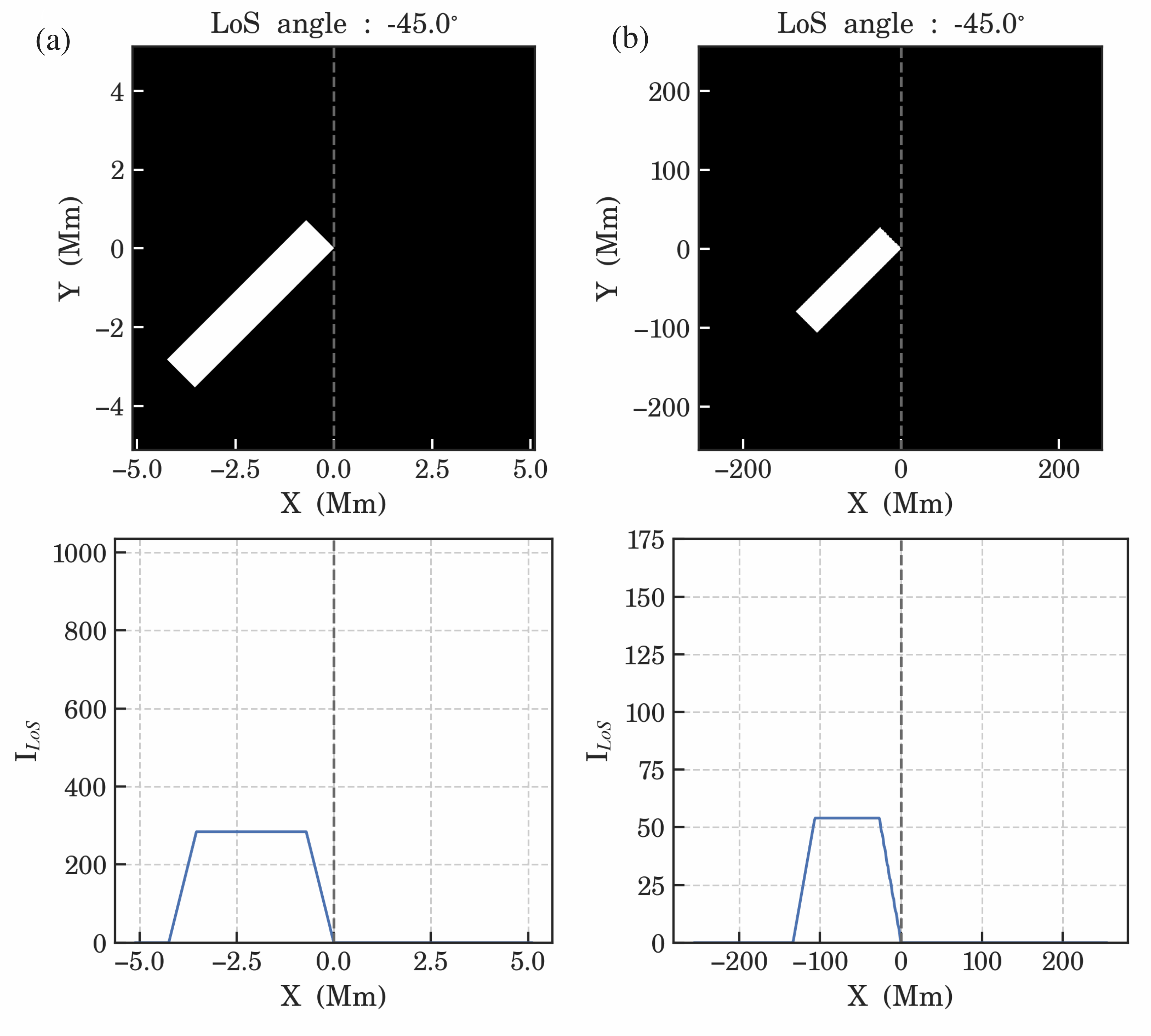}
\caption{(Top) Still example images of a toy model simulating an extended structure similar to a coronal loop, with a static point of anchoring, such that only the object's angle with respect to the LoS (denoted by the vertical dotted line) is changing, and (bottom) the LoS integrated intensity of the same structure at (a) pixel scale of 100 km/pixel. (b) pixel scale of the SDO/AIA, i.e, 435 km/pixel. This figure is available as an animation in the online version of the article. The total duration of the animation is 8 s.}
\label{fig:loseffectmodel}
\end{figure}

We have shown that the solar atmosphere, modulated by structures like plages, coronal loops, active regions, filaments etc., rotates faster than the photosphere ($\Delta\Omega\in [0.105,0.558]^{\circ}/{\rm day}$). The extended height of such features above the photosphere, especially at higher latitudes, can result in an erroneous measurement of the rotation rate based on projected coordinates \citep{Rosa1998,Vrsnak1999,Sudar2015}. Although the image correlation technique utilised in this study is tracer-independent and considers only the pixel-specific integrated intensity along the LoS to calculate the rotation rate in a particular latitudinal bin through the calculation of the 2-D cross-correlation coefficient, this method may be sensitive to the angle with respect to the LoS, and structures extending from the solar disk, such as coronal loops, whose position with respect to the LoS may influence the value of the intensity populating specific pixels and, consequently, the results obtained through image correlation. To investigate the impact of such scenarios, a simplified toy model was created, mimicking extended structures whose LoS integrated intensity changes only with respect to their position relative to the LoS, while the footpoint of the structure remains stationary. This was done to isolate the excess rotation rate resulting from the LoS effect ($\Delta\Omega_{LoS}$). The model was designed with two different spatial resolutions: (a) with a smaller pixel size corresponding to a better resolution \citep{Golub1986High}, wherein 1 pixel corresponds to 100 km in the sky; and (b) the coarser AIA pixel size, wherein 1 pixel corresponds to 435 km in the sky \citep{Alissandrakis2019} (\autoref{fig:loseffectmodel}). The aim was twofold: (i) to determine whether the LoS projection effect contributes to the disparity in rotation rate between the photosphere and the hotter solar atmosphere modulated by extended structures like coronal loops; and (ii) if it does, to assess the relevance of this effect in our study using data from SDO/AIA.

\begin{figure}
\epsscale{1.15}
    \plottwo{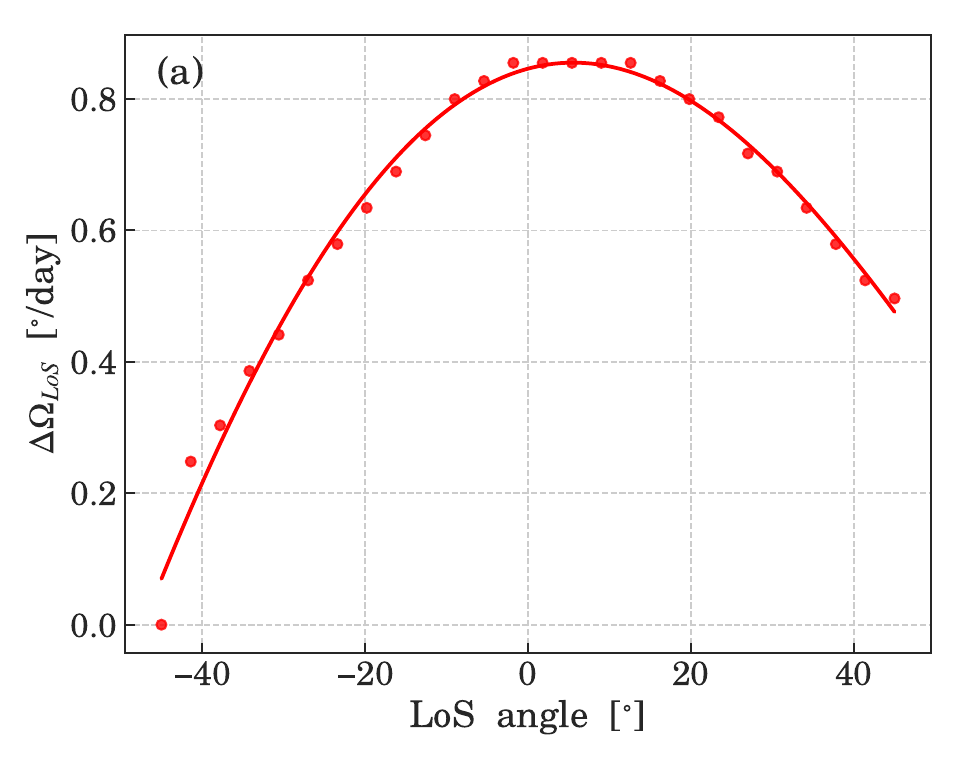}{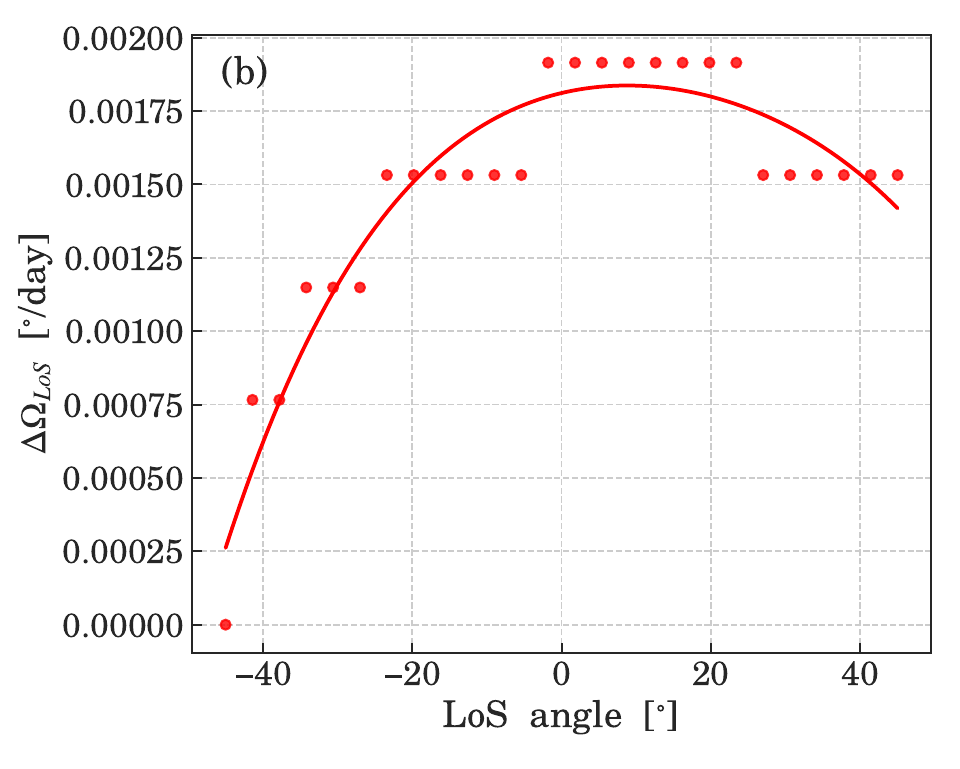}
    \caption{The variance of $\Delta\Omega_{\rm LoS}$ for change in LoS angle for a (a) 5 Mm structure in high resolution (b) 150 Mm structure in AIA resolution. An order 4 regression fit has been used to show the general trend in $\Delta\Omega_{LoS}$}
    \label{fig:loseffectshift}
\end{figure}

Once the projected intensity along the LoS (I$_{LoS}$) is obtained for a specific angle ($\theta$), the structure is shifted by a constant angle ($\Delta\theta$), which represents the anticipated change in the angle with respect to the LoS of the structure, calculated from the rotation period of the footpoint (assuming it is the solar surface) at the equator and a cadence of 6 hours. The structure is now positioned at an angle $\theta+\Delta\theta$ relative to the LoS, and the projected intensity obtained is cross-correlated with the projected intensity obtained at the initial angle $\theta$. The resulting shift is used to calculate the excess in rotation rate ($\Delta\Omega_{LoS}$), which is represented as the excess in rotation rate for the LoS angle $\theta$. This process was repeated for all LoS angles in the progression of $\theta+2\Delta\theta$, $\theta+3\Delta\theta$, and so on, spanning $\pm 45^{\circ}$ in longitude. This approach was taken to match the conditions imposed on the data in the original analysis. The results obtained for a synthetic structure of length 5 Mm observed at a high resolution of 100 km/pixel suggest an excess rotation rate ($\Delta\Omega_{LoS}$) of up to $\approx 0.83^{\circ}/{\rm day}$ (see \autoref{fig:loseffectshift} (a)), which is higher than the excess obtained in our original analysis. This suggests that when data at higher resolutions is subjected to the image correlation method without any pre-processing, there might be a LoS projection effect that affects the results obtained.

To see if such an effect can affect our analysis using AIA data, which offers a much coarser resolution, we create a much larger structure (Length$=150$\,Mm; see \autoref{fig:loseffectmodel} (b)) and subject it to the same process of analysis. The results thus obtained suggest an excess of up to $\Delta\Omega_{LoS} \approx 0.002^{\circ}/{\rm day}$ (see \autoref{fig:loseffectshift} (b)), which is not enough to explain the excess rotation rate of the extended structure-modulated solar atmosphere observed in our analysis.  In light of these results, we would like to emphasize that, as part of our analysis, we employed pre-processing techniques such as Gaussian smoothing (see \autoref{sec:Method}), which has a blurring effect and further degrades the resolution, thus minimizing the likelihood of spurious effects like the one discussed here. These findings provide further evidence that the faster rotation of the solar atmosphere is a complex physical phenomenon rather than a data or method-specific artefact.

\section{Representation of height and logarithmic temperature for AIA channels}\label{appendix:aiaheightest}
The approximate heights used to represent the different channels of the SDO/AIA are obtained from previous studies, as discussed below,
\begin{itemize}
    \item The heights and their respective uncertainties used to represent channels \threezerofour, \sixteenhundred\,  and \threethirtyfive\, are the formation heights for He {\sc ii}, C {\sc iv} and Fe {\sc xvi} emission lines as obtained in \cite{Simon1972,Simon1974,Fossum2005,Howe2012}
    \item For the channels \oneseventyone, \oneninetythree\, and \twooneone, we utilized the heights determined by \cite{Kwon2010} through the study of coronal bright points (CBPs) from the data of the \oneseventyone, $195$ {\AA}, and $284$ {\AA} channels of the Solar TErrestrial RElations Observatory (STEREO). This was done keeping in mind that the CBPs, as well, are structures which dominate in the cross-correlation process through which the rotational profile is determined for these wavelength channels of SDO/AIA. Furthermore, the logarithmic temperatures represented by the $195$ {\AA} and $284$ {\AA} channels of STEREO are nearest to the \oneninetythree\, and \twooneone\,channels of the AIA, respectively.
\end{itemize}

\begin{figure*}[!h]
\epsscale{1.15}
    \plotone{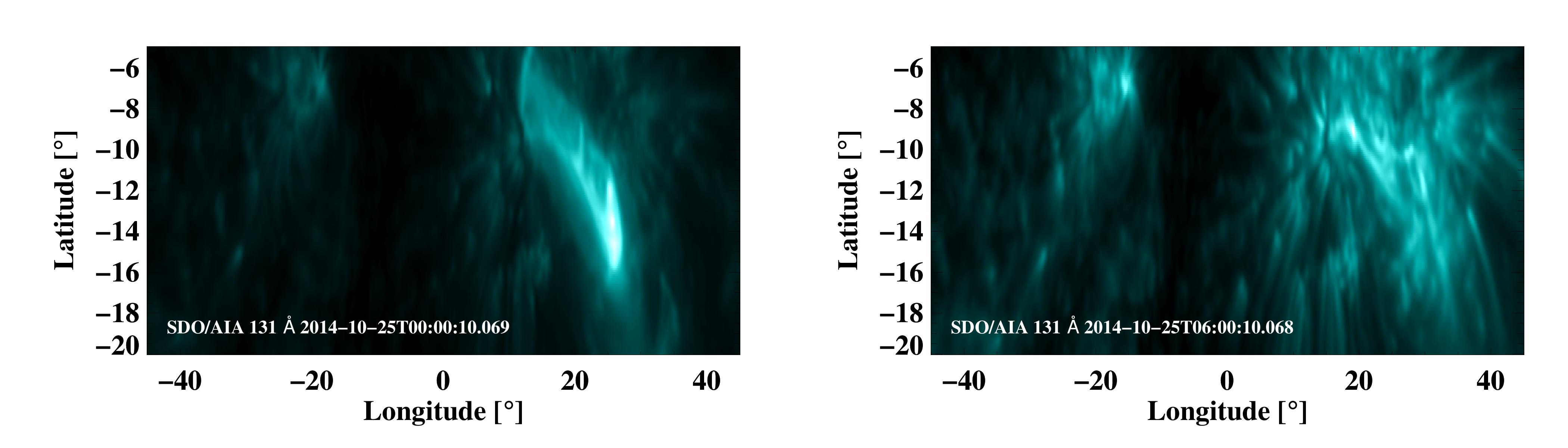}\\
    \plotone{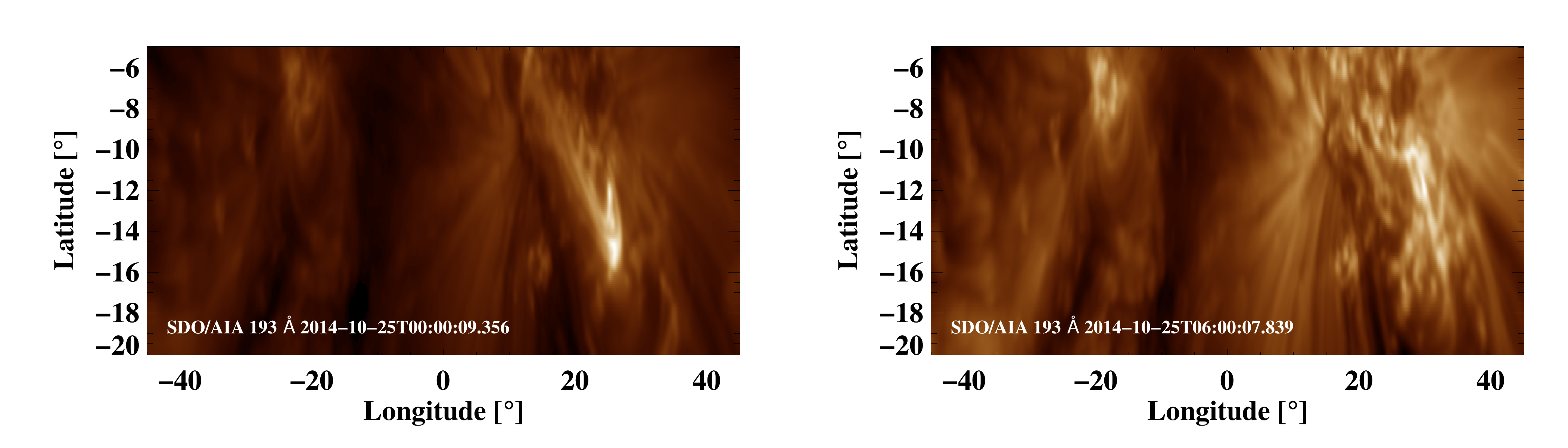}
    \caption{An example image with flare from 2014, October, 25 in heliographic coordinates after smoothing as visible in \onethirtyone\, (top) and \oneninetythree\, (bottom) channels. Since flares are spontaneous, short-term transient events with a life span much less than 6 hours, we assume they do not contribute to our long-term correlation analysis. So, the channels \onethirtyone\, and \oneninetythree\, are represented by their cooler counterparts contributed to primarily by Fe {\sc vii} and Fe {\sc xii} (${\rm Log}_{10}{\rm T} = 5.8, 6.2$), respectively.}
    \label{fig:flare131_193}
\end{figure*}

The logarithmic temperatures used to represent the wavelength channels of AIA were taken from \cite{Lemen2012,Nuevo2015} and represent the temperature responses of these respective AIA filters. Its important to point out that even though the \onethirtyone\, and \oneninetythree\, channels are also sensitive to hot flare plasma (Log$_{10}{\rm T} = 7.0$), the nature of our study focuses only on the long-term events with lifespan $> 0.25$ days or 6 hours. So, we assume the cooler component of this wavelength band contributes primarily to our results (see \autoref{fig:flare131_193}).

\bibliography{references}{}
\bibliographystyle{aasjournal}

\end{document}